# Direct Observation of k-Gaps in Dynamically Modulated Phononic Time Crystal

Ziling Liu[1, 2†], Xinghong Zhu[3†], Zhi-Guo Zhang[1], Wei-Min Zhang[1], Xue Chen[1, 2], Yong-Qiang Yang[1, 2], Ru-Wen Peng[4*], Mu Wang[4*], Jensen Li[5*], and Hong-Wei Wu[1, 2*]

[1]*School of Mechanics and Photoelectric Physics, Anhui University of Science and Technology, Huainan 232001, China*

[2]*Center for Fundamental Physics, Anhui University of Science and Technology, Huainan 232001, China*

[3]*Department of Physics, The Hong Kong University of Science and Technology, Clear Water Bay, Hong Kong, China*

[4]*National Laboratory of Solid State Microstructures, School of Physics, and Collaborative Innovation Center of Advanced Microstructures, Nanjing University, Nanjing 210093, China*

[5]*Centre for Metamaterial Research and Innovation, Department of Engineering, University of Exeter, Exeter EX4 4QF, United Kingdom*

## Abstract

Floquet time crystals, characterized by momentum gaps (k-gaps), have sparked intense interest across various branches of physics due to their intriguing dynamics and promising applications. Despite growing theoretical efforts, the realization and observation of phononic time crystals, especially for airborne sound, remain significant experimental challenges. In this work, we demonstrate a phononic time crystal by integrating discrete resonant meta-atoms into a one-dimensional acoustic waveguide, effectively creating a homogeneous, time-varying metamaterial. By dynamically modulating the effective compressibility, we experimentally observe exponential acoustic wave amplification, offering clear evidence of k-gap formation. At half the modulation frequency, we further identify an anomalous transmission peak for continuous wave excitation, whose amplitude depends on the phase difference between the incident wave and modulation cycle. Furthermore, we showcase the versatility of our platform by inducing momentum band folding and double k-gap phenomena via quasi-periodic temporal modulation. This flexible and reconfigurable approach not only enables the design of tailor-made resonant responses but also opens new avenues for realizing higher-dimensional phononic time crystals and exploring nontrivial topological dynamics in time-modulated media.

† These authors contributed equally to this work.

* Emails: hwwu@aust.edu.cn, j.li13@exeter.ac.uk, rwpeng@nju.edu.cn, muwang@nju.edu.cn

**Introduction**

Floquet time crystals are a new category of artificial materials distinct from conventional spatial crystals, whose material constitutive parameters are uniform in space but periodically modulated in time [1-5]. These periodic time interfaces induce interference between time-reflected and time-refracted waves, generating momentum band structure due to broken discrete time-translational symmetry. Unlike the energy gap ($\omega$-gap) in spatial crystals, the momentum gap (k-gap) supports two Floquet modes: one exponential growing and the other decaying in time. Recently, numerous interesting phenomena have been theoretically predicted in photonic time crystals, including topological temporal edge states [6-9], temporal Anderson localization [10], amplified emission from electrons and dipole atom [2,11], and superluminal momentum-gap solitons [12]. To realize these effects experimentally, a variety of time-varying photonic platforms have been developed. In microwave region, dynamic transmission lines are employed to experimentally observe k-gaps [13] and Bloch-Floquet and non-Bloch band structures are observed in an array of temporally driven resonators [14]. Additionally, time-varying metasurfaces have achieved exponential growth wave in k-gaps by relaxing volumetric systems into surface-based photonic time crystals [15]. However, synthesizing time-varying materials at optical frequencies remains challenging due to the need for ultrafast modulation (twice the light oscillation frequency). Promising candidates include all-optically modulated transparent conductive oxides due to the high effective of refractive index change, though high pumping power requirements lead to thermal damage and limit the performance [16-21]. Recent proposals suggest expanding k-gaps via artificial resonators with time-varying resonating strengths [22], highlighting the potential for novel modulation schemes.

As a universal concept, Floquet time crystals have also been explored in elastic wave [23-25], water wave [26], acoustics [27] and so on. For airborne sound, realizing phononic time crystals remains considerable challenge in achieving fast, spatially uniform material modulation. Prior strategies include mechanically controlled resonators for nonreciprocal transmission [28], but these suffer from frictional loss, limiting the modulation rate and depth. Electroacoustic devices with digital feedback are used to obtain nonreciprocal mode transitions [29] by temporally switching the impedance of transducers with low modulating frequencies. Furthermore, digital virtualized meta-atoms [30], which consist of microphone and speaker pairs interconnected by an external microcontroller implementing a time-varying convolution kernel, have been proposed to investigate temporal effective medium theory [31,32] at very high modulating frequencies and unidirectional amplification [33] at a low modulation frequency. These digital platforms provide great flexibility in tailoring resonant responses and modulation frequency, crucial for synthesizing homogeneous time-varying metamaterials.

In this report, we implement such phononic time crystals by integrating discrete resonating time-varying meta-atoms into a one-dimensional acoustic waveguide. By temporally modulating the effective compressibility, we experimentally observe the wave amplification, providing direct evidence of k-gap in phononic time crystals. Additionally, to highlight the platform's versatility, we also present the momentum band fold and double k-gaps opening phenomenon by quasi-periodic modulating the compressibility of metamaterial. Our experiment results demonstrate that the time-varying metamaterials composed of discrete resonating meta-atoms not only provide huge flexibility in modulating tailor-made resonating response, but also offer a promising route for constructing higher dimensional phononic time crystals and extended to other wave systems for designing Floquet time crystals.

**Results**

We begin with a one-dimensional (1D), spatially homogeneous phononic time crystal along the $x$-axis, as shown in Fig 1(a). Its compressibility $\beta(x,t)$, normalized by the compressibility of air $\beta_0$, is modulated over time ($t$) with a constant density $\rho_0$. For simplicity, we consider the temporal modulation with alternating phases A and B, over a modulation period $T_m$, shown as red and blue stripes representing $\beta_A$ and $\beta_B$. Each phase occupies $T_m/2$. For this system, the airborne acoustic wave equations are given as

$$\partial_x p(x,t) + \rho_0 \partial_t v(x,t) = 0, \tag{1}$$

$$\partial_x v + \beta_0 \partial_t (\beta(x,t) p(x,t)) = 0, \tag{2}$$

Since the medium is homogenous and infinite along the $x$ direction, the wave number $k$ remains unchanged across time interfaces. By substituting $\partial_x$ with $ik$, Eq. (1) and (2) can be written in matrix form as

$$i\partial_t \psi = \widehat{\omega}\psi, \widehat{\omega} = \begin{pmatrix} 0 & k \\ k/(\rho_0\beta) & 0 \end{pmatrix}, \tag{3}$$

with state vector $\psi = (\beta p, v)^T$. $\widehat{\omega}$ is the propagation matrix. The state vector evolves as $\psi(t) = e^{-i\widehat{\omega}t}\psi(0)$. The Floquet modes satisfies $\psi(t + T_m) = e^{-i\widehat{\omega}_B T_m/2}e^{-i\widehat{\omega}_A T_m/2}\psi(t) = e^{-i\Omega T_m}\psi(t)$, with $\widehat{\omega}_A/\widehat{\omega}_B$ being the propagation matrix in phase A/B and $\Omega$ being the Floquet frequency. This leads to the band structure ($\Omega$ verse $k$) relationship:

$$\det\left[e^{-i\Omega T_m}I_2 - e^{-i\widehat{\omega}_B T_m/2}e^{-i\widehat{\omega}_A T_m/2}\right] = 0, \tag{4}$$

where $I_2$ is the 2 by 2 identity matrix. Solving this secular equation in Eq. (4), Fig 1 (b) presents the band structure of a non-dispersive phononic crystal with compressibility modulated between $\beta_A = 1.12$ and $\beta_B = 0.88$ at a modulation frequency $1/T_m = 8.4$ kHz. A characteristic wave number associated with the modulation frequency is defined as: $k_0 = 2\pi/(c_0 T_m)$, where $c_0$ is the speed of sound in air. Then, the band structure is plotted as $\Omega T_m$ versus normalized wave number $k/k_0$, with red and blue lines denoting the real and imaginary part of Floquet frequency, respectively. The bandgap region, highlighted in yellow, near half the modulation frequency (4.2kHz), corresponds to non-zero imaginary part of Floquet frequency, one being positive (amplifying mode) and the other being negative (decay mode). To observe wave behavior around the band gap, a medium with negligible frequency dispersion is ideal and should be modulated at twice the operational frequency of interest. However, a true non-dispersive medium does not exist. As an alternative, we operate at frequencies far from the resonance of the medium to approximate a non-dispersive response. To achieve this, we design an array of acoustic resonant meta-atoms in a 1D waveguide, as shown in Fig 1 (c). The resonance parameters of each atom can be modulated in time. Each meta-atom consists of a detector and a speaker, interconnected via a microcontroller that performs time domain convolution $Y(t)$ on the detected signal, alternating between $Y_A$ (blue) and $Y_B$ (cyan). This realizes a Lorentzian-type compressibility described in the frequency domain as:

$$\beta(f) \approx 1 + \frac{c_0}{i\pi f l}Y(f), \qquad Y(f) = \frac{ifg}{f_0^2 - f^2 - 2i\gamma f}. \tag{5}$$

Here $g$, $f_0$ and $\gamma$ are the resonant strength, resonant frequency and linewidth, $l$ is the lattice distance. By modulating the resonant strength $g$ between $g_A$ and $g_B$, we effectively switch between two convolution kernels $Y_A$ and $Y_B$, realizing time-dependent compressibility. The two static compressibility configurations are shown in Fig 1 (d), labeled as "Static A" and "Static B", with the real part in red and imaginary part in blue. The compressibility values at $f_m/2$ are $\beta_A \approx 1.12$ and $\beta_B \approx 0.88$, marked with a red and blue star respectively. This approximates the nondispersive case depicted in Fig 1 (b). Using the dispersive characteristics of the two static phases, we calculate the band structure of the time modulated system plotted in Fig 1 (e), the details of calculated method are presented in Note1 of Supplementary Materials. Similar to Fig 1(b), the band structure shows a momentum gap around 4.2 kHz, which agree well with the non-dispersive case. This similarity confirms our ability to observe the momentum gap experimentally. Additionally, four more quasi energy bandgaps [34] are observed: two originate from the inherent resonance at 4.7 kHz, and the other two arise due to band folding induced by time modulation, though these are not the focus here.

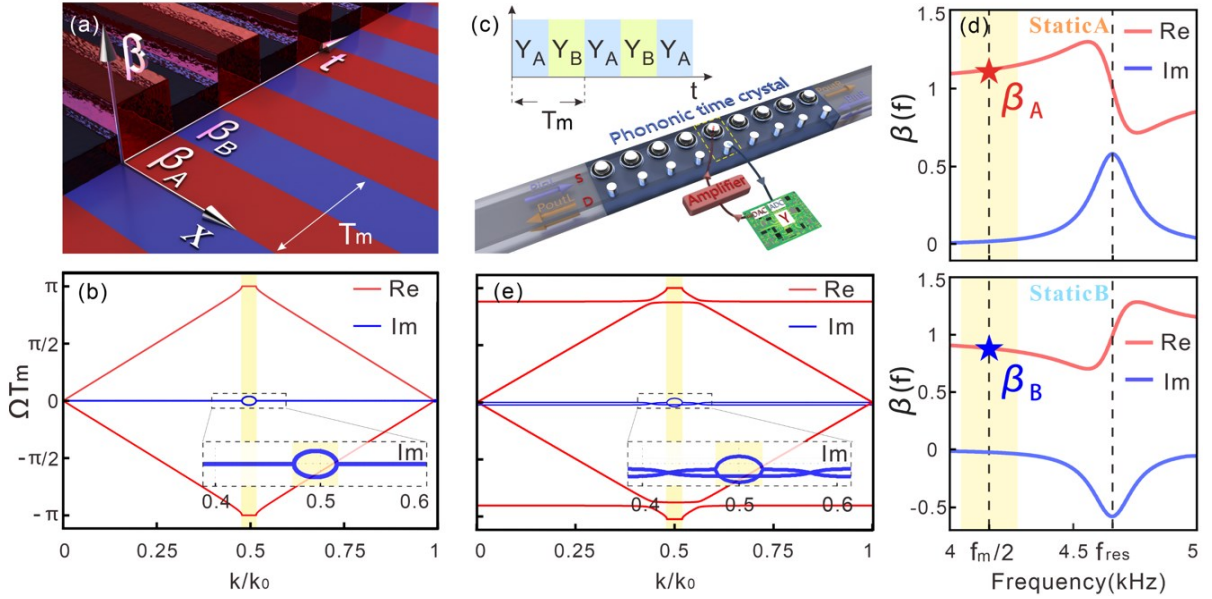

Fig 1 Concept of phononic time crystal and momentum bandgap. (a) Schematic picture of phononic time crystal whose compressibility switches between $\beta_A$ and $\beta_B$ with period $T_m$. (b) Band structure (red and blue solid lines for real and imaginary part) and momentum bandgap (orange shaded) of phononic time crystal in (a). (c) Experimental setup using a feedback system with Lorentzian resonance $Y_A$ /$Y_B$. Each meta-atom consists of a speaker and microphone interconnected by a microcontroller. The feedback is a time domain convolution switching between $Y_A$ /$Y_B$ to realize the time-varying β(t). (d) Static compressibility for phase A and B(e). (e) Band structure of phononic time crystal with time-varying β switching between phase A and B in (d).

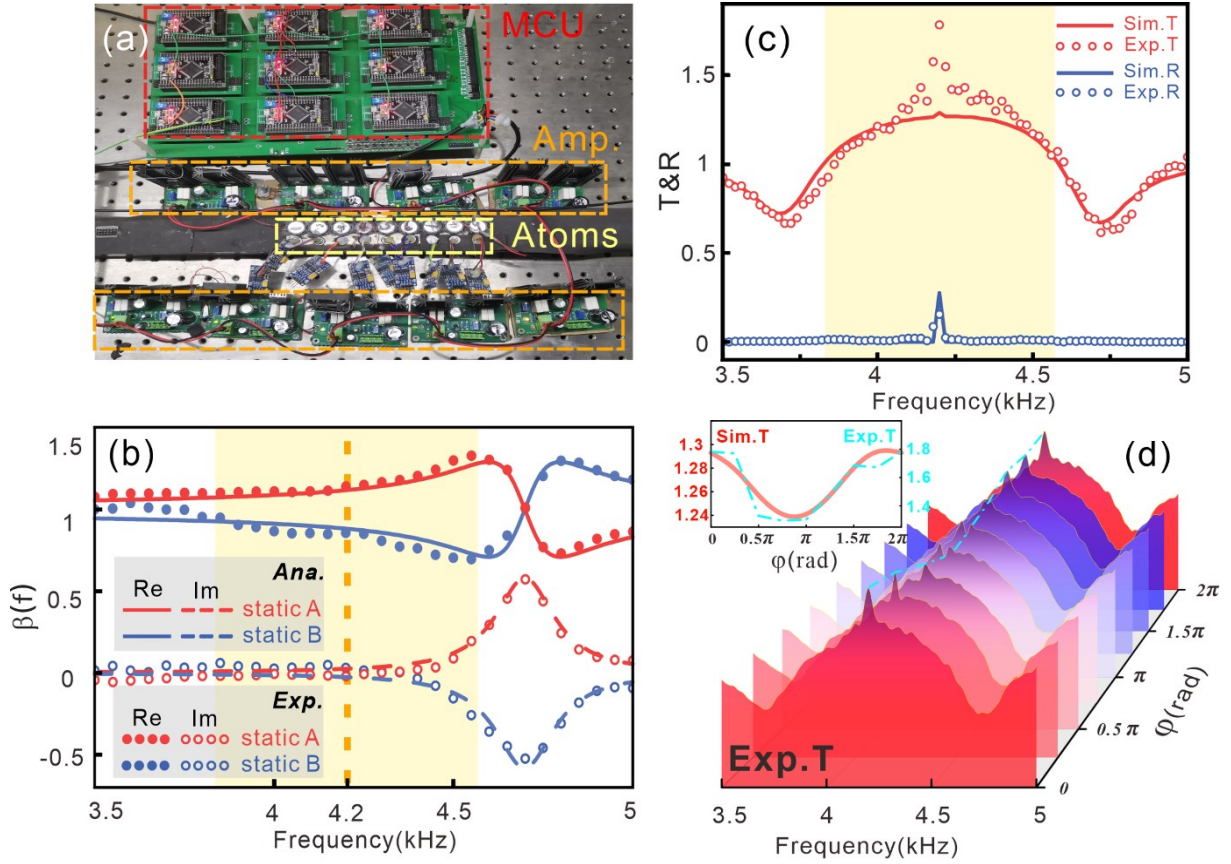

Fig 2 (a) Experimental platform of phononic time crystal with lattice distance $l = 0.02$m. (b) Static compressibility $\beta_A$ (in solid) in phase A and $\beta_B$ (in dash) in phase B with red (blue) dots representing real (imaginary) part. The parameters are chosen as $f_0 = 4.7$kHz, $g_A = 100$Hz, $g_B = -100$Hz, $\gamma = 100$Hz . (c) Experimental and simulated transmittance and reflectance for phononic time crystal with modulated compressibility between A/B phases in (b). (d) Transmission amplitude with respect to phase delay between incident wave and modulation cycle.

To observe the momentum gap shown in Fig. 1, we experimentally construct an array of 9 meta-atoms in a 1D acoustic waveguide. The experimental set up is shown in Fig 2 (a). As described earlier, each meta-atom comprises a detector and a speaker (circled with yellow dashed box), the detector senses the acoustic pressure filed in the waveguide and feeds the signal to the microcontroller (circled in red dashed box), which performs the convolution Y(t). The resulting signal is then sent to the speaker through an amplifier (circled with an orange dashed box) to generate a monopolar source. This process realizes an effective compressibility $\beta$, as described by Eq. (5). Initially, we evaluate the static compressibility of the two phases to verify the values of $\beta_A$ and $\beta_B$ in the absence of time modulation. By scanning the incident frequency, we measure the transmission $t$ and reflection $r$ coefficient and extract the effective compressibility, as shown in Fig 2 (b) with red and blue dots being the real and imaginary part. The experimental results show excellent agreement with the analytic results in Eq. (5). Notably, the imaginary part of compressibility approaches 0 within the momentum gap region.

After establishing the static configurations, we modulate between two phases and measure the transmittance $T = |t|^2$ and reflectance $R = |r|^2$ spectra, presented in red and blue dots in Fig 2 (c). In the gap region, the transmittance is larger than 1, with larger value around half of the modulation frequency, while the reflectance remains near 0. The experiment results show great agreement with the simulation results, obtained from full-wave simulation using COSMOL Multiphysics. The central peak at half the modulation frequency $f_m/2$ appears anomalous corresponding to the coupling between waves with positive and negative frequencies depended on the phase difference between the incident continuous wave and the modulation cycle. To investigate this effect, we vary the phase difference between the incident signal and the modulation cycle from 0 to $2\pi$ and remeasure the transmittance, plotted in Fig 2(d). The transmittance spectrum is largely phase insensitive at all frequencies except $f_m/2$. Accordingly, we plot the transmittance at $f_m/2$ as a function of phase delay in the inset of Fig 2(d), using bright blue dashed lines for experimental results and red lines for simulation. The transmittance goes through a cycle over $2\pi$ and a minimum occurs when the incident wave is out of phase with modulation cycle. Despite the phase variation, the transmittance consistently remains greater than 1, demonstrating the robustness of the amplifying mode within the momentum band gap. Crucially, this anomalous peak only emerges for continuous wave excitation, not sound pulses, due to the requirement for persistent phase coherence across multiple modulation periods.

In the previous section, we have experimentally demonstrated that the momentum band gap (k-gap) emerges in phononic time crystal when time modulating two static phases. In this platform, an advanced strategy involves using Lorentzian (loss) and anti-Lorentzian (gain) responses as the two phases in time modulation to enables a greater modulating depth, which results in stronger transmission and reflection within the k-gap. Furthermore, by balancing gain and loss through time domain, the effective acoustic impedance of the phononic time crystal can be engineered to approach that of air, thereby eliminating reflections at the spatial boundaries between air and the phononic time crystal. The detailed illustration is given in Note

3 of the Supplementary materials. In Fig. 3(a), we present the experimental transmittance as a function of modulating depth $\Delta a$ and frequency $f$. We observe that the transmittance increases from 1.7 to 13.4 as the modulating depth $\Delta a = \frac{2\Delta g c_0}{f_0^2 l}$ is raised from $4.8\times10^{-2}$ to $10.79\times10^{-2}$, $\Delta g$ is the resonating strength difference of each meta-atom between phases A and B. Notably, two transmission dips located at 3.7 kHz and 4.7 kHz are preserved, corresponding to the fixed resonating frequency and its lower first-order harmonic. Similarly, the reflectivity peak becomes more pronounced with increasing modulating depth in the k-gap.

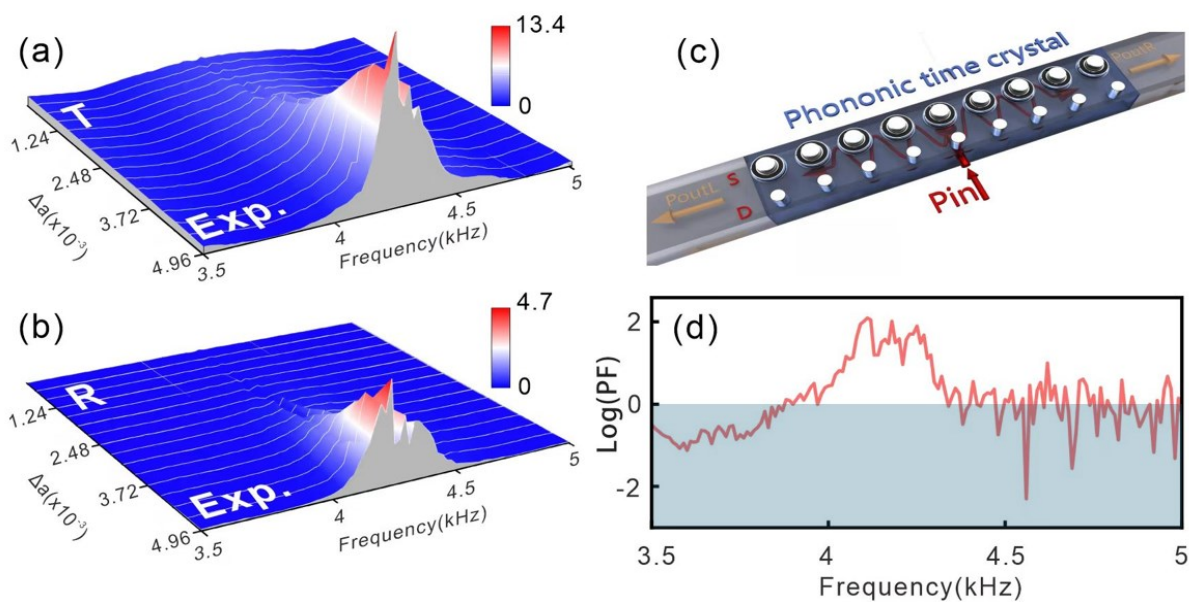

Fig 3 Transmittance (a) and reflectance (b) spectra with increased modulation depth in experiment. (c) Schematic picture for radiating source in phononic time crystal. (d) Purcell factor as a function of incident frequency.

It is well known that the amplified emission within the k-gap is a unique property of Floquet time crystals [2, 11], whose eigenmodes grow temporally over time, independent of initial phase. To demonstrate the amplified emission in phononic time crystal, we place an acoustic source at the center of the phononic time crystal structure, as shown as Fig. 3(c). To quantify the emission enhancement, we first deactivate all meta-atoms, and measure the outgoing sound pressure at both ends of the 1D empty waveguide, denoted as initial state: $p_i$. Next, we activate all meta-atoms to generate a k-gap around 4.2 kHz and record the amplified sound pressure, defining the final state: $p_f$. To character the amplified capacity of the phononic time crystal, we define the Purcell factor as $PF = \left|\frac{p_f}{p_i}\right|$, which is plotted in Fig. 3(d) as a function of incident frequency. We find that the PF can exceed $10^2$ in magnitude for a

modulating depth of $\Delta a = 10.79 \times 10^{-2}$ under finite six modulating periods. In fact, the order of amplified emission can be further enhanced by increasing either modulating depth or modulating cycles, leading to laser-like emission behavior.

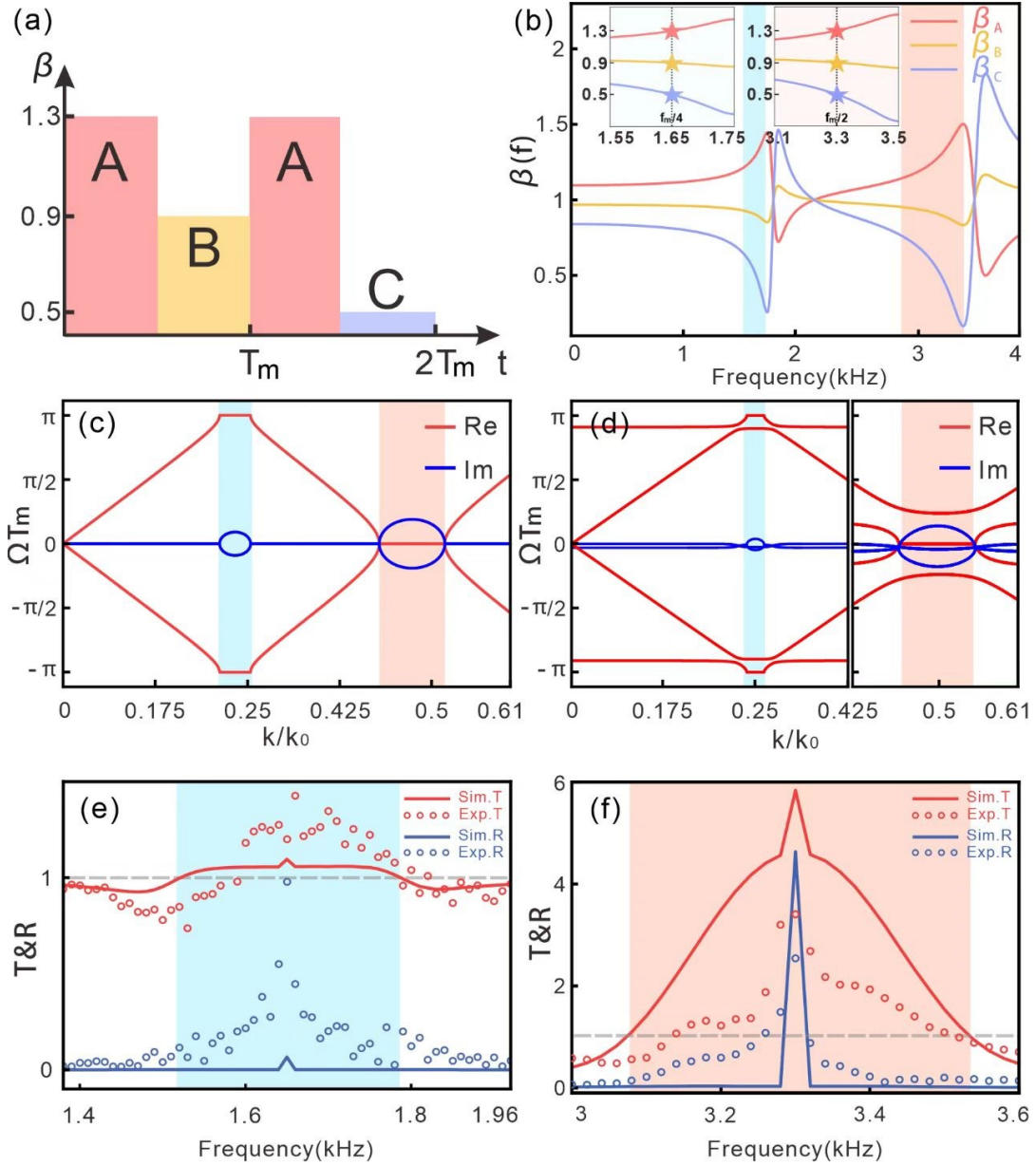

Fig 4 (a) Conceptual picture of quasi-phononic time crystal with ABAC sequence modulation. (b) Static compressibilities for different configurations to approximate nondispersive $\beta_A = 1.3$, $\beta_B = 0.9$, $\beta_C = 0.5$ around gap region. (c) Band structure of quasi-phononic time crystal in (a) with time-varying nondispersive compressibilities $\beta_A = 1.3$, $\beta_B = 0.9$ and $\beta_C = 0.5$. (d) Band structure of dispersive quasi-phononic time crystal with static phases in (b). Experimental and simulated results of transmittance and reflectance spectra around the first (e) and second k-gap (f) region.

In spatial phononic crystals, it is well known that expanding the 1D monoatomic crystal to the diatomic crystal causes the ω-band structure to fold into the first Brillouin zone and open new bandgaps [35,36]. In the same spirit of route, we construct phononic time crystal with quasi-periodic time modulation (stack-ABAC), as illustrated in Fig. 4(a). The static phases A, B, C correspond to nondispersive compressibilities $\beta_A = 1.3$, $\beta_B = 0.9$, $\beta_C = 0.5$, respectively. Each phase last for $T_m/2$. To avoid the Fabry-Pérot resonance around 2.5 kHz in the meta-atom array, we select a modulation frequency of $f_m = 6.6$ kHz as modulating frequency for this section. Thus, here the wave number correspond to the time modulation as $k_0 = \frac{2\pi f_m}{c_0} = 120.8 \ rad/m$.

According to Eq. (4), the momentum band structure can be obtained as shown in Fig. 4(c) for nondispersive phononic time crystal. Remarkably, two distinct k-gaps emerge at $k_0/2$ (light pink region) and $k_0/4$ (light cyan region), where the real part (red line) vanishes and the imaginary part (blue cycle) becomes non-zero. To experimentally observe the double k-gaps in quasi-periodical modulation, we design the system with two resonances in $Y(f)$ from Eq. (5), enabling an approximation of the desired nondispersive compressibility in both gap regions. The resulting dispersive compressibility is given by:

$$\beta(f) = 1 + \frac{a_1 f_1^2}{f_1^2 - f^2 - 2i\gamma_1 f} + \frac{a_2 f_2^2}{f_2^2 - f^2 - 2i\gamma_2 f}, \qquad (6)$$

where the resonating frequencies are $f_1 = 1.8$ kHz, $f_2 = 3.6$ kHz with decay rates $\gamma_1 = 50$Hz, $\gamma_2 = 100$Hz. Different resonating strengths can obtain different desired compressibility values. To realize $\beta_A$ at $f_m/4$ and $f_m/2$, we set $a_1 = 0.04$, $a_2 = 0.055$, as shown by the red line in Fig. 4(b). The detailed values are highlighted by the stars in the insets. The yellow Line, representing static case B corresponds to $a_1 = -0.013$, $a_2 = -0.0186$, while static case C require $a_1 = -0.067$, $a_2 = -0.092$. Figure 4(d) presents the momentum band structure of the dispersive phononic time crystal under quasi-periodic modulation. Comparing the band structures shown in Figs. 4(c) and 4(d), we observe that the dispersive phononic time crystal firmly preserves both k-gaps in $k_0/4$ and $k_0/2$ with only a slight frequency shift. Due to the high flexibility of our time-varying metamaterials in tailoring resonating responses, we experimentally implement quasi-periodic modulation for dispersive media with the compressibility shown in Fig. 4(b). Figure 4(e) shows the transmissivity and reflectivity in the first k-gap around $k_0/4$. Red lines and symbols represent simulated and experimental transmissivity, while blue lines and symbols denote the reflectivity results. The experiment results confirm the opening of the k-gap at $k_0/4$, as indicated by transmission spectrum greater than 1. A similar high transmission result is also observed at the second k-gap near $k_0/2$, as shown in Fig. 4(f). The stronger transmission in the second k-gap arises because the finite spatial length $L = 0.18$ m of phononic time crystal covers a broader range of wave numbers at higher operational frequencies, enhancing wave interaction. In brief, our versatile time-varying metamaterials successfully demonstrate the double k-gaps formation through quasi-periodic time modulation.

**Conclusion**

In this work, we have experimentally demonstrated the emergence and control of momentum band gaps (k-gaps) in phononic time crystals for airborne sound through time-periodic and quasi-periodic modulation of compressibility. Using acoustic meta-atoms with time-varying Lorentzian responses, we observed significant wave amplification within the k-gap. Building on this, we introduced quasi-periodic time modulation using three compressibility phases to realize double k-gaps at fractional modulation frequencies $k_0/4$ and $k_0/2$, supported by transmission spectral measurements. The demonstrated ability to engineer multiple, tunable k-gaps and realize laser-analog wave amplification opens new pathways for developing high-gain acoustic amplifiers, filters, and coherent sources. Our findings highlight the remarkable flexibility of time-varying acoustic metamaterials for tailoring phononic time crystals and can be extended to other Floquet metamaterials such as optics, electromagnetics, and elastic waves, promising innovative applications in wave-based technologies.

This work was supported by Scientific research projects of colleges and universities in Anhui Province (Grants No. 2022AH040114); National Key Research and Development Program of China (2022YFA1404303, 2020YFA0211300); National Natural Science Foundation of China (12234010); J L acknowledge support from the EPSRC via the META4D Programme Grant (No. EP/Y015673/1).

# Direct Observation of k-Gaps in Dynamically Modulated Phononic Time Crystal

Ziling Liu[1, 2†], Xinghong Zhu[3†], Zhi-Guo Zhang[1], Wei-Min Zhang[1], Xue Chen[1, 2], Yong-Qiang Yang[1, 2], Ru-Wen Peng[4*], Mu Wang[4*], Jensen Li[5*], and Hong-Wei Wu[1, 2*]

[1]*School of Mechanics and Photoelectric Physics, Anhui University of Science and Technology, Huainan 232001, China*

[2]*Center for Fundamental Physics, Anhui University of Science and Technology, Huainan 232001, China*

[3]*Department of Physics, The Hong Kong University of Science and Technology, Clear Water Bay, Hong Kong, China*

[4]*National Laboratory of Solid State Microstructures, School of Physics, and Collaborative Innovation Center of Advanced Microstructures, Nanjing University, Nanjing 210093, China*

[5] *Centre for Metamaterial Research and Innovation, Department of Engineering, University of Exeter, Exeter EX4 4QF, United Kingdom*

## 1. Band structure for time-varying dispersive media with a Lorentzian-type resonance

When considering the dispersion of the medium, we can write the acoustic wave equation as:

$$\partial_x p(x,t) + \rho_0 \partial_t v(x,t) = 0, \tag{S1}$$

$$\partial_x v(x,t) + \beta_0 \partial_t (p(x,t) + M(x,t)) = 0, \tag{S2}$$

with $M(x,t)$ being the monopolar polarization and governed by a Lorentzian-type model

$$\partial_t^2 M(x,t) + 2\gamma \partial_t M(x,t) + \omega_0^2 M(x,t) = a(t)\omega_0^2 p(x,t), \tag{S3}$$

Combining S1-S3 and replacing $\partial_x$ with $ik$, we can construct a 4 by 4 eigenvalue problem as

$$i\partial_t \psi = \widehat{\omega}\psi, \qquad \widehat{\omega} = \begin{pmatrix} 0 & \dfrac{k}{\beta_0} & 0 & i \\ \dfrac{k}{\rho_0} & 0 & 0 & 0 \\ 0 & 0 & 0 & -i \\ -ia(t)\omega_0^2 & 0 & i\omega_0^2 & -i2\gamma \end{pmatrix}, \qquad \text{(S4)}$$

where the state vector is $\psi = (p, v, M, -\partial_t M)^T$. The propagation matrix $\widehat{\omega}$, switches between the two phases $\widehat{\omega}_A$ and $\widehat{\omega}_B$, corresponding to the resonant strengths $a_A$ and $a_B$ in the two phases. Then, using the Floquet mode theory, we can solve the band structure plotted in Fig.1 (e) in the main text by solving the secular equation

$$\det\left[e^{-i\Omega T_m}I_4 - e^{-i\widehat{\omega}_B T_{mB}}e^{-i\widehat{\omega}_A T_{mA}}\right] = 0, \qquad \text{(S5)}$$

where $I_4$ is the 4 by 4 identity matrix.

## 2. Band structure for quasi-periodic dispersive media with two Lorentzian-type resonances

Following the same spirit of route, when we have two monopolar resonances in each phase, we only need to add another response function into the wave equation by rewriting S2 and S3 as

$$\partial_x v(x,t) + \beta_0 \partial_t(p(x,t) + M_i(x,t)) = 0, \qquad \text{(S6)}$$

$$\partial_t^2 M_i(x,t) + 2\gamma_i \partial_t M_i(x,t) + \omega_{0i}^2 M(x,t) = a_i(t)\omega_{0i}^2 p(x,t), \qquad \text{(S7)}$$

with $i = 1,2$, representing the first and second resonance mode. Combined with S1, we can construct a 6 by 6 eigenvalue problem as

$$i\partial_t\psi = \widehat{\omega}\psi, \widehat{\omega} = \begin{pmatrix} 0 & k/\beta_0 & 0 & i & 0 & i \\ k/\rho_0 & 0 & 0 & 0 & 0 & 0 \\ 0 & 0 & 0 & -i & 0 & 0 \\ -ia_1(t)\omega_{01}^2 & 0 & i\omega_{01}^2 & -2i\gamma_1 & 0 & 0 \\ 0 & 0 & 0 & 0 & 0 & -i \\ -ia_2(t)\omega_{02}^2 & 0 & 0 & 0 & i\omega_{02}^2 & -2i\gamma_2 \end{pmatrix}, \qquad \text{(S8)}$$

with $\psi = (p, v, M_1, -\partial_t M_1, M_2, -\partial_t M_2)^T$. Since now the modulation period becomes $2T_m$, and there are 3 different phases in one modulation period: A, B and C, corresponding to resonant strength pairs $(a_{1A}, a_{2A})$, $(a_{1B}, a_{2B})$ and $(a_{1C}, a_{2C})$. The modulation cycle follows the ABAC sequence, with each phase occupying $T_m/2$ giving rise to the Floquet modes: $\psi(t + 2T_m) = e^{-i\hat{\omega}_C T_m/2} e^{-i\hat{\omega}_A T_m/2} e^{-i\hat{\omega}_B T_m/2} e^{-i\hat{\omega}_A T_m/2} \psi(t)$. The band struciioooo9ikl uture can be obtained by solving the secular equation:

$$\det\left[e^{-i\Omega T_m} I_6 - e^{-i\hat{\omega}_C T_m/2} e^{-i\hat{\omega}_A T_m/2} e^{-i\hat{\omega}_B T_m/2} e^{-i\hat{\omega}_A T_m/2}\right] = 0. \qquad (S9)$$

Fig 4(d) in the main text plotted the band structure of phononic time crystal according to S9 with modulation cycle following the ABAC sequence.

## 3. Temporal effective medium for nondispersive and dispersive phononic time crystal

In main text, the nondispersive phononic time crystal is experimentally realized by dispersive time-varying metamaterial with discrete resonating meta-atoms. However, their k-gaps matching each other is limited by the low modulating depth. In fact, the temporal effective medium of nondispersive and dispersive phononic time crystal has to abide by different average methods [1]. For dispersive case, the effective compressibility can be expressed as:

$$\beta_{\text{eff}} = \xi\beta_A + (1 - \xi)\beta_B, \qquad (S10)$$

while the nondispersive case has the effective compressibility by

$$1/\beta_{\text{eff}} = \xi/\beta_A + (1 - \xi)/\beta_B, \qquad (S11)$$

where $\beta_A$ and $\beta_B$ are the static compressibilities of material A and B, $\xi$ is the duty cycle. Thus, the k-gap center $k_c$ of phononic time crystal vary with different temporal effective medium, which can be expressed as:

$$k_c = \frac{k_0}{2}\sqrt{\beta_{eff}}, \qquad\qquad (S12)$$

where wave number $k_0$ correspond to the time modulation defined as: $k_m = 2\pi/(c_0 T_m)$, where $c_0$ is the speed of sound in air. It means that the k-gap center will move with changing the effective compressibility. Thus, the k-gap center of the phononic time crystal separates for nondispersive and dispersive case due to different temporal average rules. For the dispersive phononic time crystal, balancing gain and loss through time domain, the effective compressibility always equal to 1, then $k_c = \frac{k_0}{2}$ in Fig. 1(e) and $k_c = \frac{k_0}{2}$ and $\frac{k_0}{4}$ in Fig. 4(d). Furthermore, the effective acoustic impedance of the phononic time crystal can be engineered to approach that of air, thereby eliminating reflections at the spatial boundaries between air and the phononic time crystal. However, for the nondispersive case, modulating between $\beta_A = 0.88$ and $\beta_B = 1.12$ with duty cycle $\xi = 0.5$, then the $k_c = \frac{k_0}{2} * 0.993$, thus the k-gap center slightly left shift in Fig. 1(b), this effect is more obvious for greater modulating depth in Fig. 4(c).

## 4. Stability analysis for phononic time crystals in the experiment

Since our meta-atoms involves feedback and time-varying resonant strength, the whole system (including all the atoms) is time dependent and frequency conversion occurs, we need stability analysis to make sure that the system is working in the stable regime. In this section, we develop a multiband harmonic model to determine the eigenmode (stability) of the time-dependent system. As the modulation strength and resonating frequency is modulated periodically, all the signals (from the detector and speaker) can be written in the Fourier series expansion with a set of harmonics $f_n = f + n/T_m$, where $n$ is an integer and $1/T_m$ is the modulation frequency. The detector and speaker signal can be expanded as $p(x_i, t) = \sum_n p_{in} e^{-if_n t}$, $s_i(t) = \sum_n s_{in} e^{-if_n t}$ where $i$ represents the $i$-th atom. The modulation of resonating strength and resonating frequency can be written as $g(t) = \sum_n g_n e^{-int/T_m}$, where $f_0 = 4.7$ kHz. Then, the response function Y can be written in terms of Fourier component (labeled by n harmonics): $Y_{nn'}^i = \frac{if_n g_{n-n'}}{f_0^2 - f_n^2 - 2i\gamma f_n}$ and it connects the detector and speaker signal by $s_{in} = Y_{nn'}^i p_{in'}$. Since the convolution generates the secondary radiation and propagates in waveguide and then is detected again, there is a loop transfer function from detector $D_i$ to $D_j$ by $\mathcal{T}_{\{jn\},\{in'\}}(\omega) = e^{ik_0(f_n)|x_j - x_i|} Y_{nn'}^i = e^{ik_0(f_n)|x_j - x_i|} \frac{if_n g_{n-n'}}{f_0^2 - f_n^2 - 2i\gamma f_n}$ where $|x_j - x_i|$ is the distance from speaker $i$ to detector $j$ (assuming $D_i$ and $S_i$ are in the same position since they are collocated). The $\mathcal{T}_{\{jn\},\{in'\}}$ is a square matrix and a stable system means all the poles of the system response matrix, i.e. zeros of $Det(I - \mathcal{T}_{\{jn\},\{in'\}})$ have to lie in the lower complex plane. The number of harmonics (a total of 7 in the following analysis) is taken to be enough to have converging numerical results. For the time-varying acoustic system in Fig. 2 (modulating resonant strength $g$ between $g_A = 100$Hz and $g_B = $ -100Hz, $\Delta g = 200$Hz), we plot the $Det(I - \mathcal{T}_{\{jn\},\{in'\}})$ in the complex frequency plane, as shown in Fig. S1(a). The minimum value of $Det(I - \mathcal{T}_{\{jn\},\{in'\}})$ in the blue dot indicates the pole of the system, around the frequency of $(4690-i140)$Hz, in the lower half plane. This means the system is working in the stable regime. When we increase the modulation depth, the system poles will gradually touch the real frequency axes and make the system become unstable. Fig S1 (b) plots the pole diagram where one of the poles just touch the real frequency axes, corresponding to the maximum modulation depth $\Delta g = 14.56 *$ 200Hz ($\Delta a$ =31.05×10⁻²). However, in the experiment, we only obtain the threshold at $\Delta a$ =10.79×10⁻² due to the intrinsic response of speaker and time delay of microcontroller making the system easier to become unstable.

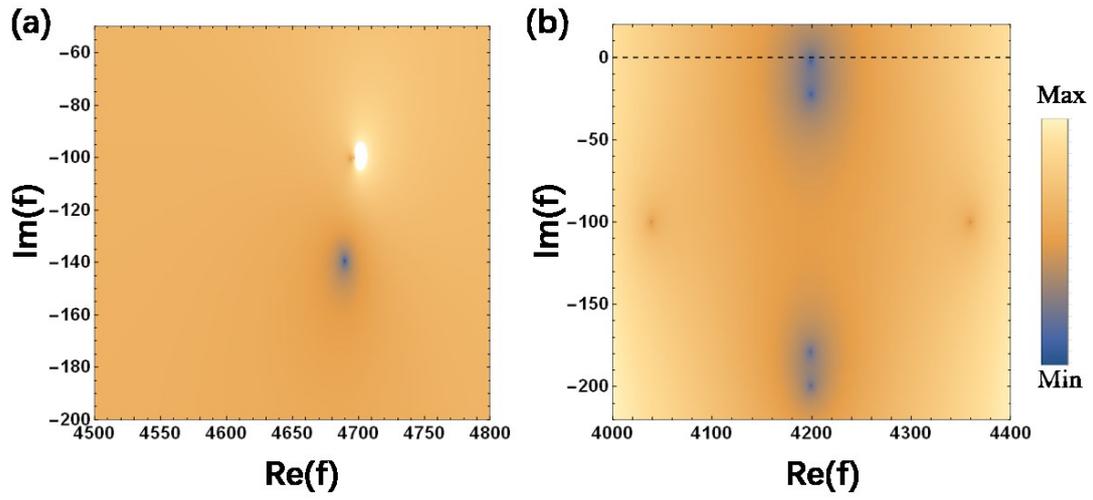

Fig S1 (a) Pole diagram of time-varying meta-atoms in Fig.2 (b) Pole diagram of time-varying meta-atoms with threshold touching the real axes.